\documentclass[twocolumn]{revtex4}
\usepackage{mathbbold}
\usepackage{amsfonts}
\usepackage{amsmath}
\usepackage{amssymb}
\usepackage{charter}
\usepackage[dvips]{graphicx}

\begin{document}

\title{Multi-headed symmetrical superpositions of coherent states}
\author{Bo Lan and Xue-xiang Xu$^{\dag }$}
\affiliation{College of Physics and Communication Electronics, Jiangxi Normal University,
Nanchang 330022, China\\
$^{\dag }$xuxuexiang@jxnu.edu.cn}

\begin{abstract}
Based on $N$ different coherent states with equal weights and phase-space
rotation symmetry, we introduce $N$-headed incoherent superposition states
(NHICSSs) and $N$-headed coherent superposition states (NHCSSs). These $N$
coherent states are associated with $N$-order roots of the same complex
number. We study and compare properties of NHICSSs\ and NHCSSs, including
average photon number, Mandel Q parameter, quadrature squeezing, Fock matrix
elements and Wigner function. Among all these states, only 2HCSS (i.e.,
Schrodinger cat state) presents quadrature-squeezing effect. Our theoretical
results can be used as a reference for researchers in this field.

\end{abstract}

\maketitle

\section{Introduction}

The superposition principle is one of the pillars upon which the entire
structure of quantum mechanics is built\cite{1}. Generally, superpositions
include operator superposition and state superposition. For example,
two-operator superposition $c_{1}\hat{O}_{1}+c_{2}\hat{O}_{2}$ is
superposing from operators $\hat{O}_{1}$\ and $\hat{O}_{2}$ with respective
weights $c_{1}$ and $c_{2}$, such as the delolated photon addition $%
a_{1}^{\dag }+e^{i\varphi }a_{2}^{\dag }$\cite{2}, ancilla-assisted photon
subtraction $a_{1}^{2}-a_{2}^{2}$\cite{3} and photon addition-subtraction
sequences $aa^{\dag }-e^{i\phi }a^{\dag }a$\cite{4}. Two-state superposition
in terms of states $\left\vert \psi _{1}\right\rangle $\ and $\left\vert
\psi _{2}\right\rangle $ devides coherent superposion $c_{1}\left\vert \psi
_{1}\right\rangle +c_{2}\left\vert \psi _{2}\right\rangle $ and incoherent
superposition $c_{1}\left\vert \psi _{1}\right\rangle \left\langle \psi
_{1}\right\vert +c_{2}\left\vert \psi _{2}\right\rangle \left\langle \psi
_{2}\right\vert $ with respective weights $c_{1}$ and $c_{2}$. Similarly,
multi-operator superposition and muliti-state superposition can be
generalized.

In quantum mechanics, the superposition principle is the origin of
nonclassical properties of quantum states\cite{5}. So superposition becomes
an important way to generating new quantum states in quantum state
engineering. Among them, coherent-state superpositions are of great
importance for many quantum subjects\cite{6}. For instance, the Schrodinger
cat state is a superposition of two coherent states with coherent amplitudes
of the same magnitude but different phases\cite{7}. Moreover, the
Schrodinger cat state can exhibit different nonclassical properties having
foundational applications in quantum information processing\cite{8,9}.

In recent several decades, the quantum superpositions involving more than
two component coherent states have been attracting increasing interests of
researchers. Yukawa et al. considered the superposition of three coherent
states with different phases\cite{10}. Raimond et al. proposed the
superposition of four coherent states with different phases\cite{11}. This
state was also called as four-headed cat state and applied to quantum phase
estimation\cite{12,13} or state preparation\cite{13a}. Vlastakis et al.
implemented the superpositions of up to four coherent states\cite{14}.

In this paper, we shall unify all these multi-component coherent-state
superpositions appearing in a large number of literatures. Unlike previous
literatures, we consider the link between $N$ coherent states in quantum
physics and $N$-order roots of complex number $\alpha $ in mathematics.
Using these coherent states, we construct superposition states with equal
weights in coherent and incoherent way. These states will present
distinctive properties. The paper is organized as follows: In Sec.II, we
introduce the background for this paper. In Sec.III, we introduce the
coherent-state superpositions with equal weights, i.e., the focus of the
paper. In Sec. IV-VI, we study and compare their statistical properties,
Fock matrix elements and Wigner functions, respectively. Conclusions are
summarized in the last section.

\section{Background}

A complex number $\alpha $ may be written in the general form $\alpha
=x+iy=re^{i\theta }$, where $x=$Re$\alpha $, $y=$Im$\alpha $, $r=\left\vert
\alpha \right\vert $, and $\theta =\arg \alpha $ are called as real part,
imaginary part, modulus and argument of $\alpha $, respectively. For a fixed
complex number $\alpha $, one can take argument $\theta =\theta _{p}+2k\pi $
for arbitrary integer $k$, where $\theta _{p}$ is the principal argument of $%
\alpha $ and varies from $0$ to $2\pi $\cite{15}. According to de Moivre
theorem, the $N$-order roots of $\alpha $ may be written as
\begin{equation}
\sqrt[N]{\alpha }=r^{\frac{1}{N}}e^{i\frac{2k\pi +\theta _{p}}{N}}\text{, }%
(k=0,1,\cdots ,N-1),  \label{1-1}
\end{equation}%
which are just $N$ complex numbers having same modulus $r^{1/N}$\ but
different phases $(2k\pi +\theta _{p})/N$. Noting that the quantity $r^{1/N}$
represents the positive $N$-order root of modulus $r$. In particular, $\sqrt[%
N]{\alpha }$\ is just $\alpha $ for $N=1$.

Geometrically, complex numbers can be shown on a so-called complex plane. In
complex plane, complex number $\alpha =x+iy$ can be drawn as a vector from $%
(0,0)$ to $(x,y)$, where one can go to $x$ point on the real axis and to $y$
point on the imaginary axis. Similarly, $N$-order roots of complex number $%
\alpha $ can be distributed symmetrically around the origin in complex
plane. Moreover, the following relation%
\begin{equation}
\sum_{k=0}^{N-1}r^{\frac{1}{N}}e^{i\frac{2k\pi +\theta _{p}}{N}}=0,
\label{1-2}
\end{equation}%
can be verified for $N\geq 2$. Indeed, they have $N$-fold rotation symmetry
and possesses $2\pi /N$ rotational symmetry, as shown in Fig.1. Recently,
discrete rotational symmetry was also used to define bosonic rotation codes%
\cite{16}.
\begin{figure}[tbp]
\label{Fig-1} \centering\includegraphics[width=0.9\columnwidth]{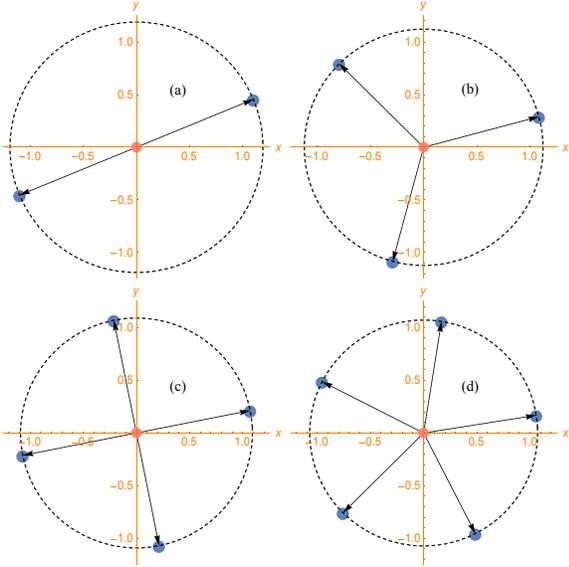}
\caption{$N$-order roots of complex number $\protect\alpha $, that is,
complex numbers $\sqrt[N]{\protect\alpha }$ in complex plane. Here we set $%
\protect\alpha $ $=1+i$ (i.e. $r=\protect\sqrt{2}$, $\protect\theta _{p}=%
\protect\pi /4$.) and (a) $N=2$; (b) $N=3$; (c) $N=4$; (d) $N=5$. Each point
in the circle denotes a root, which also corresponds a vector from the
origin to the point.}
\end{figure}

As we all know, coherent state $\left\vert \alpha \right\rangle $ with
amplitude $\alpha $\ can be generated by operating the displacement operator
$D\left( \alpha \right) =e^{\alpha a^{\dag }-\alpha ^{\ast }a}$\ on the
vacuum $\left\vert 0\right\rangle $\cite{17,18}. That is to say, one to one
correspondence can be established between a coherent state and a complex
number. Therefore, we can obtain $N$ different coherent states $\left\vert
r^{\frac{1}{N}}e^{i\frac{2k\pi +\theta _{p}}{N}}\right\rangle $ in terms of
above $N$-order roots of $\alpha $.

\section{Coherent-state superpositions with equal weights}

Employing equal-weight superpositions of above $N$ coherent states, we
introduce multi-headed quantum states in this section.

\textit{Superposition way I:} A $N$-headed incoherent superpostion state (%
\textit{NHICSS})%
\begin{equation}
\rho _{ic}=\frac{1}{N}\sum_{k=0}^{N-1}\left\vert r^{\frac{1}{N}}e^{i\frac{%
2k\pi +\theta _{p}}{N}}\right\rangle \left\langle r^{\frac{1}{N}}e^{i\frac{%
2k\pi +\theta _{p}}{N}}\right\vert .  \label{2-1}
\end{equation}%
is introduced by superposing above $N$ coherent states with equal weights in
incoherent way. Obviously, this state is the equal-weight incoherent mixture
of those $N$ coherent states. Recently, this state have been used as the
resource of quantum key distribution\cite{19,20}.

\textit{Superposition way II:} A $N$-headed coherent superposition state (%
\textit{NHCSS})
\begin{equation}
\left\vert \psi _{c}\right\rangle =\frac{1}{\sqrt{N_{c}}}\sum_{k=0}^{N-1}%
\left\vert r^{\frac{1}{N}}e^{i\frac{2k\pi +\theta _{p}}{N}}\right\rangle ,
\label{2-2}
\end{equation}%
is introduced by superposing above $N$ coherent states with equal weights in
coherent way. Its density operator, i.e. $\rho _{c}=\left\vert \psi
_{c}\right\rangle \left\langle \psi _{c}\right\vert $, can be further
written as
\begin{equation}
\rho _{c}=\frac{1}{N_{c}}\sum_{k_{1}=0}^{N-1}\sum_{k_{2}=0}^{N-1}\left\vert
r^{\frac{1}{N}}e^{i\frac{2k_{1}\pi +\theta _{p}}{N}}\right\rangle
\left\langle r^{\frac{1}{N}}e^{i\frac{2k_{2}\pi +\theta _{p}}{N}}\right\vert
.  \label{2-3}
\end{equation}%
Here
\begin{equation}
N_{c}=\sum_{k_{1}=0}^{N-1}\sum_{k_{2}=0}^{N-1}e^{r^{\frac{2}{N}}(e^{2\pi i%
\frac{k_{1}-k_{2}}{N}}-1)}.  \label{2-4}
\end{equation}%
is the normalization factor. In particularly, $N_{c}=1$ for case $N=1$ and $%
N_{c}=2+2e^{-2r}$ for case $N=2$. Moreover, for case $N=1$, $\rho _{c}$ and $%
\rho _{ic}$\ will reduce to the coherent state $\left\vert \alpha
\right\rangle $. By the way, the NHCSS $\left\vert \psi _{c}\right\rangle $
is the eigenstate of the operator $a^{N}$.

In contrast to coherent state with $N=1$, we call states with $N\geq 2$ as
multi-head cases. In the field of quantum optics, the 2HCSS\ is often called
as the Schrodinger cat state\cite{21}. Following this naming rule, NHCSSs
(i.e. high-order cat states) are called the generalized cat states\cite{22}.
Different from other literatures, we introduce NHICSSs\ and NHCSSs
associated with $N$-order roots of the same $\alpha $. For example, the
Schrodinger cat state is not defined as $\left\vert \alpha \right\rangle
+\left\vert -\alpha \right\rangle $ but as $\left\vert \sqrt{\left\vert
\alpha \right\vert }\right\rangle +\left\vert -\sqrt{\left\vert \alpha
\right\vert }\right\rangle $\ (2HCSS) in our work.

\section{Statistical properties}

In order to calculate statistical properties, we firstly derive $%
\left\langle a^{\dag h}a^{l}\right\rangle $ ($h,l$ are intergers) for
NHICSSs\ and NHCSSs. For NHICSS, we have the general expression%
\begin{equation}
\left\langle a^{\dag h}a^{l}\right\rangle _{\rho _{ic}}=\frac{r^{\frac{h+l}{N%
}}}{N}\sum_{k=0}^{N-1}e^{il\frac{2k\pi +\theta _{p}}{N}-ih\frac{2k\pi
+\theta _{p}}{N}}.  \label{3-1}
\end{equation}%
Using Eq.(\ref{3-1}), we list $\left\langle a^{\dag }\right\rangle $, $%
\left\langle a\right\rangle $, $\left\langle a^{\dag }a\right\rangle $, $%
\left\langle a^{\dag 2}\right\rangle $, $\left\langle a^{2}\right\rangle $,
and $\left\langle a^{\dag 2}a^{2}\right\rangle $ for NHICSS in table I.
Moreover, we find $\left\langle a^{\dag }a\right\rangle =\left\vert \alpha
\right\vert ^{2/N}$ and $\left\langle a^{\dag 2}a^{2}\right\rangle
=\left\vert \alpha \right\vert ^{4/N}$ for NHICSS in table I. For NHCSS, we
have the general expression%
\begin{eqnarray}
\left\langle a^{\dag h}a^{l}\right\rangle _{\rho _{c}} &=&\frac{r^{\frac{h+l%
}{N}}}{N_{c}}\sum_{k_{1}=0}^{N-1}\sum_{k_{2}=0}^{N-1}e^{r^{\frac{2}{N}%
}(e^{2\pi i\frac{k_{1}-k_{2}}{N}}-1)}  \notag \\
&&\times e^{il\frac{2k_{1}\pi +\theta _{p}}{N}-ih\frac{2k_{2}\pi +\theta _{p}%
}{N}},  \label{3-2}
\end{eqnarray}%
Using Eq(\ref{3-2}), we list $\left\langle a^{\dag }\right\rangle $, $%
\left\langle a\right\rangle $, $\left\langle a^{\dag }a\right\rangle $, $%
\left\langle a^{\dag 2}\right\rangle $, $\left\langle a^{2}\right\rangle $,
and $\left\langle a^{\dag 2}a^{2}\right\rangle $ for NHCSS in table II. Some
expressions of $\left\langle a^{\dag }a\right\rangle $\ and $\left\langle
a^{\dag 2}a^{2}\right\rangle $\ in Table II, which are related to $%
\left\vert \alpha \right\vert $ but not $\theta _{p}$, haven't been given
yet because of their complexity. But we can easily provide their numerical
results through the computing software of \textbf{Mathematica}.
Interestingly, we we have $\left\langle a^{\dag }\right\rangle =\left\langle
a\right\rangle =\left\langle a^{\dag 2}\right\rangle $ $=\left\langle
a^{2}\right\rangle =0$ for cases $N\geq 3$ in both table I and table II. \
\begin{table}[h]
\caption{Expectation values $\left\langle a^{\dag h}a^{l}\right\rangle $ for
some NHICSSs with same $\protect\alpha $.}
\begin{center}
\begin{tabular}{|c||c|c|c|c|c|c|}
\hline\hline
Case & $\left\langle a^{\dag }\right\rangle $ & $\left\langle a\right\rangle
$ & $\left\langle a^{\dag }a\right\rangle $ & $\left\langle a^{\dag
2}\right\rangle $ & $\left\langle a^{2}\right\rangle $ & $\left\langle
a^{\dag 2}a^{2}\right\rangle $ \\ \hline\hline
$N=1$ & $\alpha ^{\ast }$ & $\alpha $ & $\left\vert \alpha \right\vert ^{2}$
& $\alpha ^{\ast 2}$ & $\alpha ^{2}$ & $\left\vert \alpha \right\vert ^{4}$
\\
$N=2$ & $0$ & $0$ & $\left\vert \alpha \right\vert $ & $\alpha ^{\ast }$ & $%
\alpha $ & $\left\vert \alpha \right\vert ^{2}$ \\
$N=3$ & $0$ & $0$ & $\left\vert \alpha \right\vert ^{2/3}$ & $0$ & $0$ & $%
\left\vert \alpha \right\vert ^{4/3}$ \\
$N=4$ & $0$ & $0$ & $\left\vert \alpha \right\vert ^{1/2}$ & $0$ & $0$ & $%
\left\vert \alpha \right\vert $ \\
$N=5$ & $0$ & $0$ & $\left\vert \alpha \right\vert ^{2/5}$ & $0$ & $0$ & $%
\left\vert \alpha \right\vert ^{4/5}$ \\
$N=6$ & $0$ & $0$ & $\left\vert \alpha \right\vert ^{1/3}$ & $0$ & $0$ & $%
\left\vert \alpha \right\vert ^{2/3}$ \\
$\vdots $ & $\vdots $ & $\vdots $ & $\vdots $ & $\vdots $ & $\vdots $ & $%
\vdots $ \\ \hline
\end{tabular}%
\end{center}
\end{table}
\begin{table}[h]
\caption{Expectation values $\left\langle a^{\dag h}a^{l}\right\rangle $ for
some NHCSSs with same $\protect\alpha $.}
\begin{center}
\begin{tabular}{|c||c|c|c|c|c|c|}
\hline\hline
Case & $\left\langle a^{\dag }\right\rangle $ & $\left\langle a\right\rangle
$ & $\left\langle a^{\dag }a\right\rangle $ & $\left\langle a^{\dag
2}\right\rangle $ & $\left\langle a^{2}\right\rangle $ & $\left\langle
a^{\dag 2}a^{2}\right\rangle $ \\ \hline\hline
$N=1$ & $\alpha ^{\ast }$ & $\alpha $ & $\left\vert \alpha \right\vert ^{2}$
& $\alpha ^{\ast 2}$ & $\alpha ^{2}$ & $\left\vert \alpha \right\vert ^{4}$
\\
$N=2$ & $0$ & $0$ & $\left\vert \alpha \right\vert \tanh \left\vert \alpha
\right\vert $ & $\alpha ^{\ast }$ & $\alpha $ & $\left\vert \alpha
\right\vert ^{2}$ \\
$N=3$ & $0$ & $0$ & $\left( \cdots \right) $ & $0$ & $0$ & $\left( \cdots
\right) $ \\
$N=4$ & $0$ & $0$ & $\left( \cdots \right) $ & $0$ & $0$ & $\left( \cdots
\right) $ \\
$N=5$ & $0$ & $0$ & $\left( \cdots \right) $ & $0$ & $0$ & $\left( \cdots
\right) $ \\
$N=6$ & $0$ & $0$ & $\left( \cdots \right) $ & $0$ & $0$ & $\left( \cdots
\right) $ \\
$\vdots $ & $\vdots $ & $\vdots $ & $\vdots $ & $\vdots $ & $\vdots $ & $%
\vdots $ \\ \hline
\end{tabular}%
\end{center}
\end{table}

\textit{Average photon number:} Light intensity can be described by average
photon number $\bar{n}=\left\langle \hat{n}\right\rangle =\left\langle
a^{\dagger }a\right\rangle $. In Fig.2, we plot $\bar{n}$ as a function of $%
\left\vert \alpha \right\vert $ for NHICSSs and NHCSSs. As shown in Fig.2(a)
for NHICSS, in the regime of $0<\left\vert \alpha \right\vert <1$, the
larger $N$ is, the larger $\bar{n}$ is; while in the regime of $\left\vert
\alpha \right\vert >1$, the larger $N$ is, the smaller $\bar{n}$ is. This
can also be checked by $\bar{n}=\left\vert \alpha \right\vert ^{2/N}$. While
from Fig.2 (b) for NHCSS, in the whole range of $\left\vert \alpha
\right\vert $, the larger $N$ is, the smaller $\bar{n}$ is.
\begin{figure}[tbp]
\label{Fig-2} \centering\includegraphics[width=0.9\columnwidth]{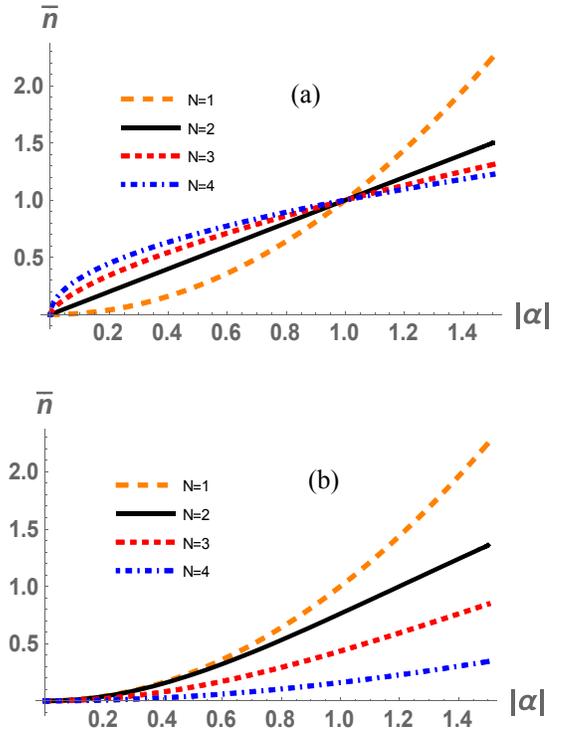}
\caption{$\bar{n}$ versus $\left\vert \protect\alpha \right\vert $ for (a)
NHICSSs and (b) NHCSSs.}
\end{figure}

\textit{Mandel Q parameter:} We examine the Mandel Q parameter $%
M_{Q}=\left\langle a^{\dagger 2}a^{2}\right\rangle /\left\langle a^{\dagger
}a\right\rangle -\left\langle a^{\dagger }a\right\rangle $, which indicate
that the distribution is Poissonian, super-Poissonian and sub-Poissonian, if
$M_{Q}=0$, $M_{Q}>0$ or $M_{Q}<0$, respectively. Obviously, coherent state
and NHICSSs are Poissonian, which can be verified by $M_{Q}=\left\vert
\alpha \right\vert ^{4}/\left\vert \alpha \right\vert ^{2}-\left\vert \alpha
\right\vert ^{2}=0$ and $M_{Q}=\left\vert \alpha \right\vert
^{4/N}/\left\vert \alpha \right\vert ^{2/N}-\left\vert \alpha \right\vert
^{2/N}=0$. While for NHCSSs, we plot $M_{Q}$ as a function of $\left\vert
\alpha \right\vert $ in Fig.3. For 2HCSS, the distribution is
super-Poissonian when $\left\vert \alpha \right\vert <11.7069$ and
Poissionian when $\left\vert \alpha \right\vert >11.7069$. For 3HCSS,\ the
distribution is super-Poissionian when $\left\vert \alpha \right\vert
<5.23972$, sub-Poissionian when $5.23972<\left\vert \alpha \right\vert
<17.1512$, and Poissionian when $\left\vert \alpha \right\vert >17.1512$.
For 4HCSS,\ the distribution is super-Poissionian when $\left\vert \alpha
\right\vert <9.8696$ and $39.4784<\left\vert \alpha \right\vert <88.8264$,
sub-Poissionian when $9.8696<\left\vert \alpha \right\vert <88.8264$, and
Poissionian when $\left\vert \alpha \right\vert >88.8264$. Moreover, all
these states will tend to Poisson distribution if $\left\vert \alpha
\right\vert $ is large enough. \
\begin{figure}[tbp]
\label{Fig-3} \centering\includegraphics[width=0.9\columnwidth]{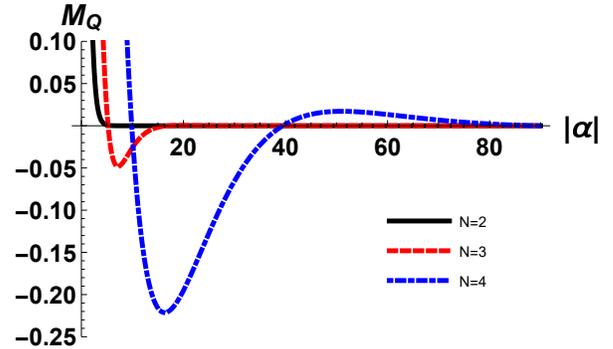}
\caption{$M_{Q}$ versus $\left\vert \protect\alpha \right\vert $ for NHCSS
with $N=2$, $N=3$, and $N=4$.}
\end{figure}

\textit{Quadrature squeezing effect:} We explore squeezing of quadrature
amplitude defining from quadratures $X_{1}=(a+a^{\dag })/\sqrt{2}\ $and $%
X_{2}=(a-a^{\dag })/(\sqrt{2}i)$. Their variances can be expressed as $%
\left\langle \Delta X_{j}\right\rangle ^{2}=\left\langle a^{\dag
}a\right\rangle -\left\vert \left\langle a^{\dag }\right\rangle \right\vert
^{2}\pm $ Re$(\left\langle a^{\dag 2}\right\rangle -\left\langle a^{\dag
}\right\rangle ^{2})+0.5$ with $+\rightarrow j=1$ and $-\rightarrow j=2$,
respectively. A quantum state exhibits quadrature squeezing if $\Delta
^{2}X_{1}<0.5$ or $\Delta ^{2}X_{2}<0.5$. In fact, we know $\Delta
^{2}X_{1}=0.5$ and $\Delta ^{2}X_{2}=0.5$ for coherent (vacuum) state.
Obviously, NHCSSs and NHICSSs with $N\geq 3$ have no squeezing due to $%
\left\langle \Delta X_{j}\right\rangle ^{2}=\left\langle a^{\dag
}a\right\rangle +0.5\geq 0$. 2HICSS has no squeezing due to $\left\langle
\Delta X_{j}\right\rangle ^{2}=\left\vert \alpha \right\vert \pm $ Re$%
(\alpha ^{\ast })+0.5\geq 0.5$. But 2HCSS has the possibility of squeezing
in certain range of $\left\vert \alpha \right\vert $, which can be seen from
Fig.4 and $\left\langle \Delta X_{j}\right\rangle ^{2}=\left\vert \alpha
\right\vert \tanh \left\vert \alpha \right\vert \pm $ Re$(\alpha ^{\ast
})+0.5$.\qquad\
\begin{figure}[tbp]
\label{Fig-4} \centering\includegraphics[width=0.9\columnwidth]{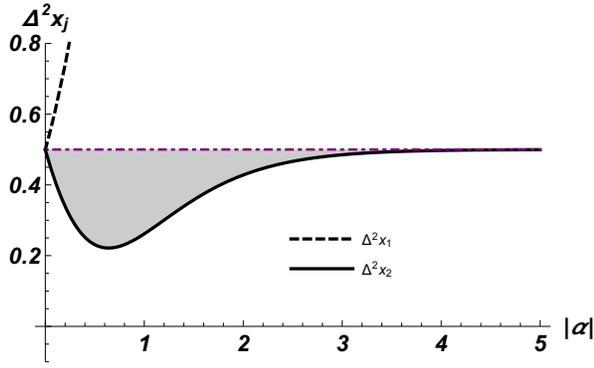}
\caption{$\left\langle \Delta X_{j}\right\rangle ^{2}$ as a function of $%
\left\vert \protect\alpha \right\vert $ for 2HCSS. The squeezing effect is
shown in the grey region.}
\end{figure}

\section{Fock matrix elements}

A density operator $\rho $ can be written as $\rho =\sum_{m=0}^{\infty
}\sum_{n=0}^{\infty }p_{mn}\left\vert m\right\rangle \left\langle
n\right\vert $ with $p_{mn}=\left\langle m\right\vert \rho \left\vert
n\right\rangle $. Here, the coefficients $p_{mn}$ are called Fock matrix
elements (FME) corresponding to component $\left\vert m\right\rangle
\left\langle n\right\vert $\cite{23,23a}. Of course, the coefficients $%
p_{mm} $ show the photon number distribution (PND).

FMEs of NHICSS can be expressed as%
\begin{equation}
p_{mn}^{ic}=\frac{r^{\frac{m+n}{N}}e^{-r^{\frac{2}{N}}}}{N\sqrt{m!n!}}%
\sum_{k=0}^{N-1}e^{im\frac{2k\pi +\theta _{p}}{N}-in\frac{2k\pi +\theta _{p}%
}{N}},  \label{4-1}
\end{equation}%
PNDs for NHICSS can be written as $p_{mm}^{ic}=r^{2m/N}e^{-r^{2/N}}/m!$ (a
Poissionian distribution).

FMEs of NHCSS can be expressed as%
\begin{equation}
p_{mn}^{c}=\frac{r^{\frac{m+n}{N}}e^{-r^{\frac{2}{N}}}}{N_{c}\sqrt{m!n!}}%
\sum_{k_{1}=0}^{N-1}\sum_{k_{2}=0}^{N-1}e^{im\frac{2k_{1}\pi +\theta _{p}}{N}%
-in\frac{2k_{2}\pi +\theta _{p}}{N}},  \label{4-2}
\end{equation}%
From Eq.(\ref{2-2}), NHCSS can expanded in the photon-number basis%
\begin{equation}
\left\vert \psi _{c}\right\rangle =\frac{Ne^{-r^{2/N}/2}}{\sqrt{N_{c}}}%
\sum_{s=0}^{\infty }\frac{\alpha ^{s}}{\sqrt{\left( N\cdot s\right) !}}%
\left\vert N\cdot s\right\rangle .  \label{4-3}
\end{equation}%
so PNDs for NHCSS can be written as $p_{mm}^{c}=\left\vert \left\langle
m\right\vert \left. \psi _{c}\right\rangle \right\vert
^{2}=N^{2}r^{2s}e^{-r^{2/N}}\delta _{m,N\cdot s}/\left( N_{c}\left( N\cdot
s\right) !\right) $.

In order to see the distributions, we take $N=3$ as example and plot FMEs in
Fig.5 and PNDs in Fig.6. Obviously, 3HICSS contains populations\ and
coherences of all Fock states $\left\vert 0\right\rangle $, $\left\vert
1\right\rangle $, $\left\vert 2\right\rangle $, $\cdots $. But for 3HCSS, it
contains populations\ and coherences of Fock states $\left\vert
0\right\rangle $, $\left\vert 3\right\rangle $, $\left\vert 6\right\rangle $%
, $\cdots $. Moreover, 3HCSS with $\alpha =1+i$\ is close to the
superposition $\sqrt{3/4}\left\vert 0\right\rangle +\sqrt{1/4}\left\vert
3\right\rangle $, which is just the superposition of Fock states $\left\vert
0\right\rangle $\ and $\left\vert 3\right\rangle $. This similar state has
been generated in experiment\cite{10}.

\begin{figure}[tbp]
\label{Fig-5} \centering\includegraphics[width=0.9\columnwidth]{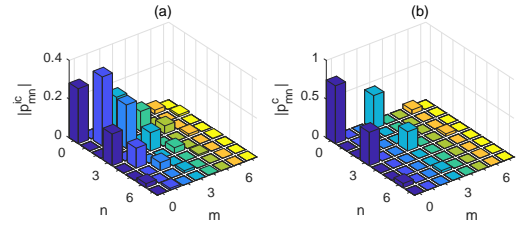}
\caption{FMEs for (a) 3HICSS and (b) 3HCSS with $\protect\alpha =1+i$.}
\end{figure}

\begin{figure}[tbp]
\label{Fig-6} \centering\includegraphics[width=0.9\columnwidth]{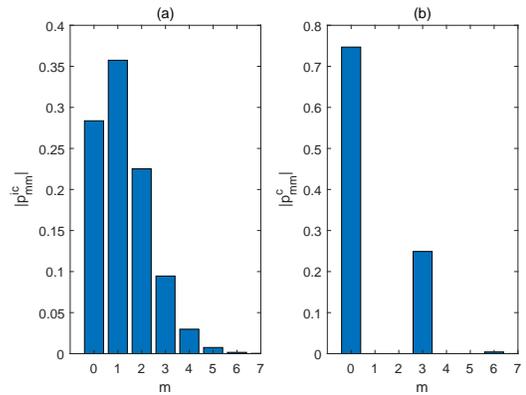}
\caption{PNDs for (a) 3HICSS and (b) 3HCSS with $\protect\alpha =1+i$.}
\end{figure}

\section{Wigner function}

Wigner function is extremely useful in quantum optics because it contains
complete information about quantum state\cite{24}. Wigner function of a
quantum state $\rho $ can be expressed as $W_{\rho }\left( \beta \right) =%
\frac{2}{\pi }\mathrm{Tr}[\rho D\left( \beta \right) \Pi D^{\dag }\left(
\beta \right) ]$ with complex-number coordinate $\beta =\left( x+iy\right) /%
\sqrt{2}$ in phase space, where $\Pi =\left( -1\right) ^{a^{\dag }a}$
denotes the photon number parity operator and $D\left( \beta \right)
=e^{\beta a^{\dag }-\beta ^{\ast }a}$ is the displacement operator\cite{14}.
In normal-order form, $W_{\rho }\left( \beta \right) $ can be written as\cite%
{25}%
\begin{equation}
W_{\rho }\left( \beta \right) =\left\langle \frac{2}{\pi }:e^{-2\left(
a^{\dag }-\beta ^{\ast }\right) \left( a-\beta \right) }:\right\rangle
_{\rho }.  \label{5-1}
\end{equation}%
For NHICSS, we have%
\begin{equation}
W_{\rho _{ic}}\left( \beta \right) =\frac{2}{\pi N}\sum_{k=0}^{N-1}e^{-2%
\left\vert r^{\frac{1}{N}}e^{i\frac{2k\pi +\theta _{p}}{N}}-\beta
\right\vert ^{2}}.  \label{5-2}
\end{equation}%
For NHCSS, we have%
\begin{eqnarray}
W_{\rho _{c}}\left( \beta \right) &=&\frac{2}{\pi N_{c}}\sum_{k_{1}=0}^{N-1}%
\sum_{k_{2}=0}^{N-1}e^{r^{\frac{2}{N}}(e^{2\pi i\frac{k_{1}-k_{2}}{N}}-1)}
\notag \\
&&\times e^{-2(r^{\frac{1}{N}}e^{-i\frac{2k_{2}\pi +\theta _{p}}{N}}-\beta
^{\ast })(r^{\frac{1}{N}}e^{i\frac{2k_{1}\pi +\theta _{p}}{N}}-\beta )}.
\label{5-3}
\end{eqnarray}

In particularly, for case $N=1$, Eqs.(\ref{5-2}) and (\ref{5-3}) will reduce
to Wigner function of coherent state $\left\vert \alpha \right\rangle $
\begin{equation}
W_{\left\vert \alpha \right\rangle }\left( \beta \right) =\frac{2}{\pi }%
e^{-2\left\vert \alpha -\beta \right\vert ^{2}},  \label{5-4}
\end{equation}%
which have a Gaussian form with the center $\alpha $\ in phase space. For
case $N=2$, Eq.(\ref{5-2}) will reduce to%
\begin{equation}
W_{\rho _{ic}}^{N=2}\left( \beta \right) =\frac{e^{-2\left\vert \sqrt{r}%
e^{i\theta _{p}/2}-\beta \right\vert ^{2}}}{\pi }+\frac{e^{-2\left\vert -%
\sqrt{r}e^{i\theta _{p}/2}-\beta \right\vert ^{2}}}{\pi }.  \label{5-5}
\end{equation}%
Similarly, Eq.(\ref{5-3}) will reduce to%
\begin{eqnarray}
W_{\rho _{c}}^{N=2}\left( \beta \right) &=&\frac{e^{-2\left\vert \sqrt{r}%
e^{i\theta _{p}/2}-\beta \right\vert ^{2}}}{\pi \left( 1+e^{-2r}\right) }+%
\frac{e^{-2\left\vert -\sqrt{r}e^{i\theta _{p}/2}-\beta \right\vert ^{2}}}{%
\pi \left( 1+e^{-2r}\right) }  \notag \\
&&+\frac{e^{-2\sqrt{r}e^{-i\theta _{p}/2}\beta +2\sqrt{r}e^{i\theta
_{p}/2}\beta ^{\ast }-2\left\vert \beta \right\vert ^{2}}}{\pi \left(
1+e^{-2r}\right) }  \notag \\
&&+\frac{e^{+2\sqrt{r}e^{-i\theta _{p}/2}\beta -2\sqrt{r}e^{i\theta
_{p}/2}\beta ^{\ast }-2\left\vert \beta \right\vert ^{2}}}{\pi \left(
1+e^{-2r}\right) },  \label{5-6}
\end{eqnarray}%
where the last two terms owe to the interference between $\left\vert \sqrt{%
\left\vert \alpha \right\vert }\right\rangle $ and $\left\vert -\sqrt{%
\left\vert \alpha \right\vert }\right\rangle $.

Using Eqs.(\ref{5-4}), (\ref{5-5}) and (\ref{5-6}), we depict Wigner
functions for coherent state, 2HICSS, 2HCSS, respectively in Fig.7. As
represented pictorially in Fig.7(a), the Wigner function of coherent state $%
\left\vert \alpha \right\rangle $\ with $\alpha =\left\vert \alpha
\right\vert e^{i\theta _{p}}$ have a positive Gaussian peak, whose center of
shaded circle (representing \textquotedblleft area of
uncertainty\textquotedblright ) locates at distance $\left\vert \alpha
\right\vert $\ from the origin and at angle $\theta _{p}$\ above the
position axis. From Fig.7(b), Wigner function of 2HICSS has a characteristic
shape that consists of two positive Gaussian peaks. From the Fig.7(c), we
can see that Wigner function of 2HCSS has more hills and dips\cite{26}.\
\begin{figure}[tbp]
\label{Fig-7} \centering\includegraphics[width=0.9\columnwidth]{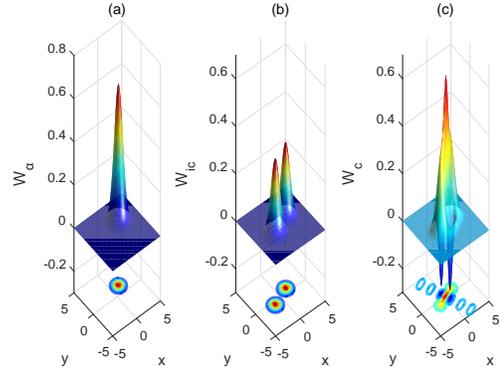}
\caption{Wigner functions for (a) coherent state $\left\vert \protect\alpha %
\right\rangle $; (b) 2HICSS; (c) 2HCSS. with $\protect\alpha =1+i$.}
\end{figure}
\begin{figure}[tbp]
\label{Fig-8} \centering\includegraphics[width=0.9\columnwidth]{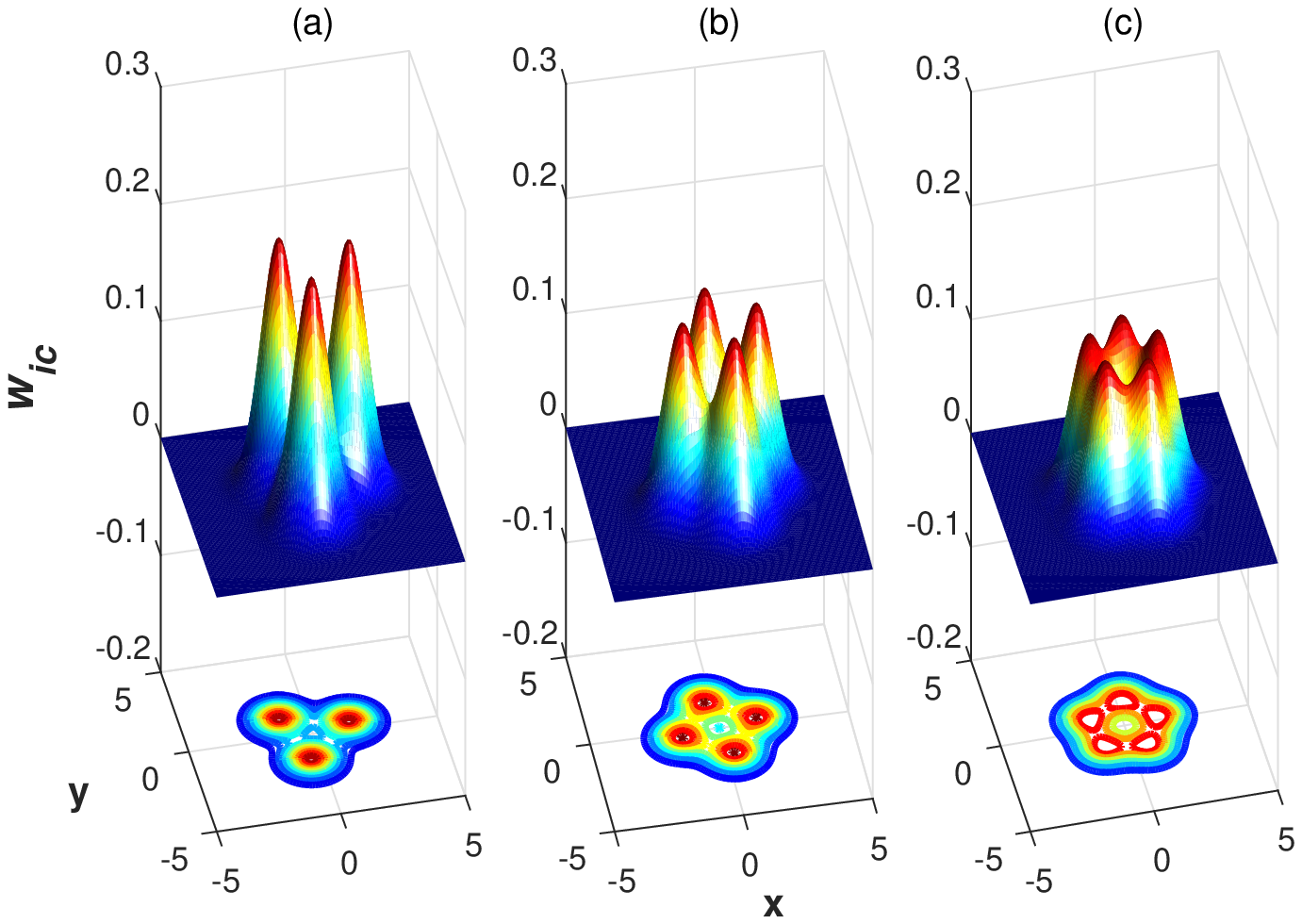}
\caption{Wigner functions for (a) 3HICSS; (b) 4HICSS; (c) 5HICSS with $%
\protect\alpha =1+i$.}
\end{figure}
\begin{figure}[tbp]
\label{Fig-9} \centering\includegraphics[width=0.9\columnwidth]{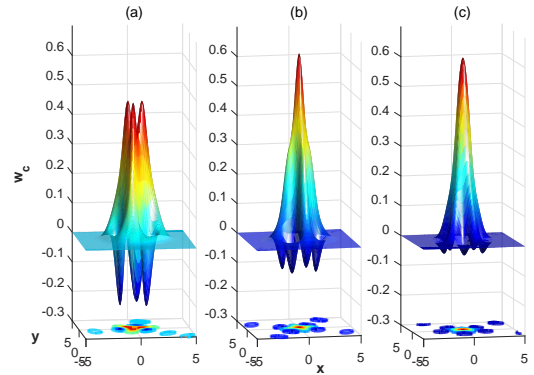}
\caption{Wigner functions for (a) 3HCSS; (b) 4HCSS; (c) 5HCSS with $\protect%
\alpha =1+i$.}
\end{figure}
\qquad \qquad \qquad \qquad \qquad \qquad \qquad \qquad\ \qquad

According to Eq.(\ref{5-2}), we plot Wigner functions for 3HICSS, 4HICSS and
5HICSS in Fig.8. Clearly, the distribution of peaks is same as the
distribution of $N$ roots of $\alpha $ in complex plane. According to Eq.(%
\ref{5-3}), we plot Wigner functions for 3HCSS, 4HCSS and 5HCSS in Fig.9.
Compared with Fig.8, surfaces in Fig.9 have remarkably exhibit\ multiple
areas of negativity in phase space, which indicate the nonclassicality of
the NHCSSs. Indeed, multiple areas of negativity result from higher
interference effects of coherent states. Of course, Wigner functions in both
Fig.8 and Fig.9 possess distinctive rotational $2\pi /N$\ symmetry.

Meanwhile, we can get the parity values through the relation $\left\langle
\Pi \right\rangle =\frac{\pi }{2}W_{\rho }\left( 0\right) $ after knowing
the Wigner function\cite{27}. In Fig.10, we depict the parity value $%
\left\langle \Pi \right\rangle $\ of different NHICSSs and NHCSSs as a
function of $\left\vert \alpha \right\vert $. For NHICSSs in Fig.10 (a), we
find that: (1) in the regime of $0<\left\vert \alpha \right\vert <1$, the
larger $N$ is, the smaller $\left\langle \Pi \right\rangle $ is; (2) in the
regime of $\left\vert \alpha \right\vert >1$, the larger $N$ is, the larger $%
\left\langle \Pi \right\rangle $ is; (3) Furthermore, $\left\langle \Pi
\right\rangle $ decreases with the increase of $\left\vert \alpha
\right\vert $ and finally tends to zero; (4) A fixed value $\left\langle \Pi
\right\rangle =1/e^{2}\simeq 0.135335$ will find for different NHICSSs\ if $%
\left\vert \alpha \right\vert =1$. For NHCSSs in Fig.10 (b), it can be seen
that: (1) $\left\langle \Pi \right\rangle $ of even NHCSSs is always 1,
which means that even NHCSSs only contain even photon number components; (2)
$\left\langle \Pi \right\rangle $ of odd NHCSSs will vary in the range of -1
to 1; (3) $\left\langle \Pi \right\rangle $ of odd NHCSSs will be limit to $%
0 $ for large $\left\vert \alpha \right\vert $.
\begin{figure}[tbp]
\label{Fig-10} \centering\includegraphics[width=0.9\columnwidth]{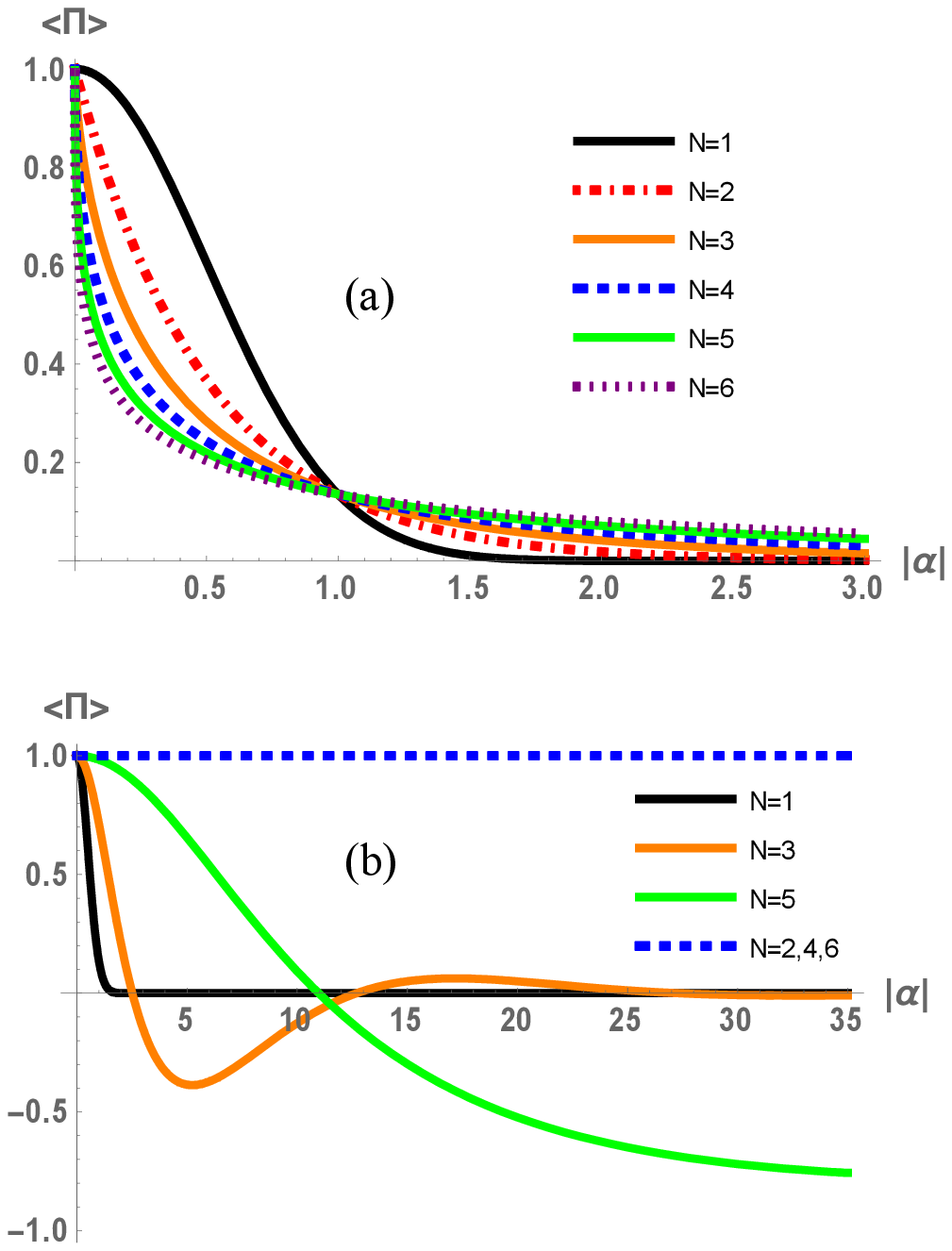}
\caption{Parity $\left\langle \Pi \right\rangle $ vesus $\left\vert \protect%
\alpha \right\vert $ for (a) NHICSSs and (b) NHCSSs.}
\end{figure}
\qquad

\section{Conclusions and discussions}

To conclude, we have introduced NHICSSs and NHCSSs by superposing $N$
coherent states associated with $N$-order roots of complex number $\alpha $.
Therefore, our research belongs to mathematical physics. In fact, these
quantum states have also been widely studied in previous literatures. But
different from those literatures, we have unified all the states in the
standard form based on the same complex number. Some properties, including
average photon number, Mandel Q parameter, quadrature squeezing effect, Fock
matrix elements (photon number distributions) and Wigner function, have been
studied for these quantum states in detail. Anlytical expressions have been
given and numerical results have been analyzed.

Our main results show that: (1) Light intensity of NHCSS will decrease as $N$
increases for any $\left\vert \alpha \right\vert $; but that of NHICSS will
increase ($\left\vert \alpha \right\vert <1$) and decrease ($\left\vert
\alpha \right\vert >1$) as $N$ increases. (2) NHICSS remains the Poissionian
character of the original coherent state, but NHCSS may present Poissionian,
sub-poissionian and super-Poissionian in different $\left\vert \alpha
\right\vert $. (3) Only 2HCSS may present the quadrature squeezing effect.
(4) NHICSSs include all photon components, but NHCSSs only include photon
components with $\left\vert N\cdot s\right\rangle $; (5) Wigner functions of
NHICSSs have no negative region, but Wigner functions of NHCSSs have
negative regions due to the interference effect. In addition, the parity has
been discussed incidentally.

By the way, we have only considered superpositions of coherent states with
the equal weights. In principle, many other coherent-state superpositions
will generate by setting arbitrary superposition weights. Moreover, quantum
states related to our considered NHICSSs\ and NHCSSs are useful for quantum
metrology \cite{28,29,30}, quantum error correction\cite{31,32} and quantum
key distribution\cite{33,34,35}. We believe that these theoretical results
will provide further references for relevant researchers.

\textbf{Disclosure statement:} No potential conflict of interest was
reported by the authors.

\textbf{Funding:} This project was supported by the National Natural Science
Foundation of China (Grant number: 11665013).


\begin{thebibliography}{99}
\bibitem{1} P. A. Dirac, \textit{The principle of quantum mechanics}
(Cambridge University Press, Cambridge, 1930).

\bibitem{2} N. Biagi, L. S. Costanzo, M. Bellini, and A. Zavatta, Phys. Rev.
Lett. \textbf{124}, 033604 (2020).

\bibitem{3} H. Takahashi, K. Wakui, S. Suzuki, M. Takeoka, K. Hayasaka, A.
Furusawa, and M. Sasaki, Phys. Rev. Lett. \textbf{101}, 233605 (2008).

\bibitem{4} A. Zavatta, V. Parigi, M. S. Kim, H. Jeong, and M. Bellini,
Phys. Rev. Lett. \textbf{103}, 140406 (2009).

\bibitem{5} F. R. Cardoso, D. Z. Rossatto, G. P. L. M. Fernandes, G.
Higgins, and C. J. Villas-Boas, Phys. Rev. A \textbf{103}, 062405 (2021).

\bibitem{6} N. Akhtar, B. C. Sanders, and C. Navarrete-Benlloch, Phys. Rev.
A \textbf{103}, 053711 (2021).

\bibitem{7} E. Schrodinger, Naturwissenschaften \textbf{23}, 807 (1935).

\bibitem{8} G. Tatsi, L. Mazzarella, and J. Jeffers, Phys. Rev. A \textbf{103%
}, 023709 (2021).

\bibitem{9} K. K. Mishra, D. Yadav, G. Shukla, and D. K. Mishra, Phys. Scr.
\textbf{96}, 045102 (2021).

\bibitem{10} M. Yukawa, K. Miyata, T. Mizuta, H. Yonezawa, P. Marek, R.
Filip, and A. Furusawa, Opt. Express \textbf{21}, 5529 (2013).

\bibitem{11} J. M. Raimond, C. Sayrin, S. Gleyzes, I. Dotsenko, M. Brune, S.
Haroche, P. Facchi, and S. Pascazio, Phys. Rev. Lett. \textbf{105}, 213601
(2010).

\bibitem{12} S. Y. Lee, C. W. Lee, H. Nha, and D. Kaszlikowski, J. Opt. Soc.
Am. \textbf{32}, 061186 (2015).

\bibitem{13} D. A. R. Dalvit, R. L. de Matos Filho, and F. Toscano, New J.
Phys. \textbf{8}, 010276 (2006).

\bibitem{13a} L. Y. Jiang, Q. Guo, X. X. Xu, M. Cai, W. Yuan, and Z. L.
Duan, Opt. Commun. \textbf{369}, 179 (2016).

\bibitem{14} B. Vlastakis, G. Kirchmair, Z. Leghtas, S. E. Nigg, L. Frunzio,
S. M. Girvin, M. Mirrahimi, M. H. Devoret, and R. J. Schoelkopf, Science
\textbf{342}, 607 (2013).

\bibitem{15} C. Harper, \textit{Analytic Methods in Physics} (Wiley-VCH,
Berlin, 1999).

\bibitem{16} A. L. Grimsmo, J. Combes, and B. Q. Baragiola, Phys. Rev. X
\textbf{10}, 011058 (2020).

\bibitem{17} R. J. Glauber, Phys. Rev. \textbf{131}, 2766 (1963).

\bibitem{18} M. O. Scully and M. S. Zubairy, Quantum Optics (Cambridge
University Press, 1997).

\bibitem{19} A. Denys, P. Brown, and A. Leverrier, Quantum \textbf{5}, 540
(2021).

\bibitem{20} P. Papanastasiou, C. Lupo, C. Weedbrook, and S. Pirandola,
Phys. Rev. A \textbf{98}, 012340 (2018).

\bibitem{21} C. C. Gerry and P.L. Knight, \textit{Introductory Quantum Optics%
} (Cambridge University Press, 2005).

\bibitem{22} A. Z. Goldberg and K. Heshami, arXiv: 2106.03862 (2021).

\bibitem{23} H. L. Zhang, H. C. Yuan, and X. X. Xu, Phys. Scr. \textbf{95},
045101 (2020).

\bibitem{23a} G. Pierobon, G. Cariolaro, and G. Dattoli, J. Math. Phys.
\textbf{62}, 082101 (2021).

\bibitem{24} W. P. Schleich, \textit{Quantum Optics in Phase Splace}
(WILEY-VCH Verlag GmbH, Berlin, 2001).

\bibitem{25} Z. M. Liu and L. Zhou, Optik \textbf{142}, 1 (2017).

\bibitem{26} D. V. Sychev, A. E. Ulanov, A. A. Pushkina, M. W. Richards, I.
A. Fedorov, and A. I. Lvovsky, Nature Photon. \textbf{57}, 379 (2017).

\bibitem{27} R. J. Birrittella, P. M. Alsing, and C. C. Gerry, AVS Quantum
Sci. \textbf{3}, 014701 (2021).

\bibitem{28} A. Z. Goldberg, A. B. Klimov, M. Grassl, G. Leuchs, and L. L.
Sanchez-Soso, AVS Quantum Sci. \textbf{2}, 044701 (2020);

\bibitem{29} D. A. R. Dalvit, R. L. de Matos Filho, and F. Toscano, New J.
Phys. \textbf{8}, 276 (2006).

\bibitem{30} N. Akhtar, B. C. Sanders, and C. Navarrete-Benlloch, Phys. Rev.
A \textbf{103}, 053711 (2021).

\bibitem{31} M. Bergmann, and P. van Loock, Phys. Rev. A \textbf{94}, 042332
(2016).

\bibitem{32} A. L. Grimsmo, J. Combes, and B. Q. Baragiola, Phys. Rev. X
\textbf{10}, 011058 (2020).

\bibitem{33} D. Sych and G. Leuchs, New J. Phys. \textbf{12}, 053019 (2010).

\bibitem{34} P. Papanastasiou and S. Pirandola, Phys. Rev. Res. \textbf{3},
013047 (2021).

\bibitem{35} J. Lin, T. Upadhyaya, and N. Lutkenhaus, Phys. Rev. X \textbf{9}%
, 041064 (2019).
\end{thebibliography}
\end{document}